\documentclass[aps,pre,twocolumn,groupedaddress,longbibliography,showkeys,showpacs]{revtex4-1}

\usepackage{graphicx}
\usepackage{bm}
\usepackage{amssymb,amsfonts}
\usepackage{amsmath}
\usepackage{amsthm}
\usepackage{natbib}
\usepackage{graphicx}
\usepackage[usenames,dvipsnames]{xcolor}
\usepackage{soul}
\usepackage{url}
\graphicspath{{Figures/}}

\begin{document}

\title{On Clustering Coefficients in Complex Networks}


\author{Alexander I Nesterov}
   \email{nesterov@academicos.udg.mx}
\affiliation{Departamento de F{\'\i}sica, CUCEI, Universidad de Guadalajara,
 Guadalajara, CP 44430, Jalisco, M\'exico}

\date{\today}

\begin{abstract}
The clustering coefficient is a valuable tool for understanding the structure of complex networks. It is widely used to analyze social networks, biological networks, and other complex systems. While there is generally a single common definition for the local clustering coefficient, there are two different ways to calculate the global clustering coefficient. The first approach takes the average of the local clustering coefficients for each node in the network. The second one is based on the ratio of closed triplets to all triplets. It is shown that these two definitions of the global clustering coefficients are strongly inequivalent and may significantly impact the accuracy of the outcome.
\end{abstract}

 \keywords{complex networks; statistical mechanics; graph ensembles; clustering coefficient; hidden variables;  graph temperature}

\maketitle

In network science, a measure called the clustering coefficient shows how many nodes in a network tend to group. It provides insight into the local cohesiveness of connections in a network. A high clustering coefficient indicates a network with a community structure where nodes form tightly interconnected groups. In contrast, a low clustering coefficient implies a more random or decentralized network structure \cite{DUJW,DSMF, BABM, GCal, BAL1, BAL, ARB,KJSD, MN2018,GMNM,VIHP,WDSS,BIGI,CCFSLV,BAGN,SSTH,GMNM,MEJN1,Watts1302,NMEJ1,FORTUNATO,ROZXX}.

 A local clustering coefficient (LCC) of a node $i$ is the ratio of the number of connections between the neighbors of the node to the total possible connections among them: \begin{align}
c_i=\frac{2 \times \text {number of triangles centered on node } \mathrm{i}}{\text { degree of } i \cdot(\text { degree of } i-1)}
\end{align}
The global clustering coefficient (GCC) measures the tendency for nodes in a graph to cluster together. The GCC ranges from 0 to 1, with 0 indicating that the graph is entirely unclustered, while one suggests that every node is part of a closed triangle \cite{WDSS,NMSW,BSLV,ARB}. 

There are two possible definitions of the GCC \cite{BAWM,PJNM}. The first one is the definition of the GCC as the average of the local clustering coefficients of all nodes in the network:
\begin{align}
	C_1 =\frac{1}{N}\sum_i c_i ,
	  \label{EqC1}
\end{align}
where $N$ is the number of nodes and $c_i$ is the LCC of the node $i$. The second definition is as follows:
\begin{align}
    C_2=\frac{3\times\text { number of triangles }}{\text { number of connected triples }}.
    \label{EqC2}
    \end{align}  
The definitions for the GCCs introduced above are non-equivalent, i.e., in some situations, one can obtain $C_1=1$ and $C_2=0$ (see, for instance, the discussion in  Refs. \cite{BBOR,WYGEV}).
  
  In this Letter, we show that these two definitions of the GCCs are strongly inequivalent. We consider the fermionic exponential random graph model (ERGM) with hidden variables to prove this claim. We show that the coefficient $C_2$ yields an expected behavior for the lower-temperature regime, while $C_1$ gives an ``incorrect'' answer. 

{\em Model.} -- We consider an undirected fermionic graph with a fixed number of nodes and a varying number of links. The model belongs to the class of ERGMs and is extensively explained in \cite{NAVPH}. The connection probability of the link between nodes $i$ and $j$ is given by
\begin{align}
	 p_{i j} =\frac{1}{e^{\beta \left(\varepsilon_{i j}-\mu\right) }+ 1},
	 \label{EqP}
\end{align}
where $\beta = 1/T$ is the inverse temperature of the network, $\mu$ is the chemical potential;  $\varepsilon_{i j} = \varepsilon_{i } + \varepsilon_{j}$ and $\varepsilon_i $ is an ``energy"   assigned to each node $i$. It is supposed that $0 \leq \varepsilon_i  \leq \mu$. 

The expected degree of a node $i$ is given by $\bar k_i= \sum_j p_{ij}$. Denoting the average node degree in the whole network with $\langle  k \rangle = (1/N)\sum_i  \bar k_i $, we obtain
\begin{align}
\langle  k \rangle  =\frac{2}{N}\sum _{i < j } \frac{1}{e^{\beta \left(\varepsilon_{i j}-\mu\right) }  +1}.
\label{Eq.4_1}
    \end{align} 
For $N \gg 1$ one can replace the sums by integrals:  $\frac{1}{(N-1)}\sum_{i} \rightarrow \int$ and $\frac{2}{N(N-1)}\sum_{i<j} \rightarrow \iint$. In the continuous limit, the expected degree of a node with energy $\varepsilon$ and  the average node degree in the whole network can be recast as
\begin{align}
	& \bar k (\varepsilon ) = (N - 1)\int p(\varepsilon, \varepsilon' ) \rho(\varepsilon' ) d \varepsilon' ,
	 \label{EqEK1} \\
&\langle  k \rangle  = (N -1)  \iint  p(\varepsilon, \varepsilon')  \rho (\varepsilon ) \rho (\varepsilon')  d \varepsilon d \varepsilon',
\label{Eq.4}
    \end{align}
 where
\begin{align}
    p(\varepsilon, \varepsilon') = \frac{1}{e^{\beta \left(\varepsilon+ \varepsilon' -\mu\right) }+1},
    \label{FD}
\end{align}
and $\rho (\varepsilon)$ denotes the density of states given by
\begin{align}\label{Eq5a}
    \rho(\varepsilon) =&\frac{\alpha\beta e^{\alpha \beta (\varepsilon- 
\mu/2)}}{2\sinh(a\beta\mu/2)}.
\end{align}
Here  $\alpha = \beta_c (\gamma -1)/\beta$, $0 \leq \varepsilon \leq \mu$, and the standard normalization condition, $\int_0^{\mu }\rho (\varepsilon) d \varepsilon =1$ is imposed.

The expected node degree and the average node degree per node, $ \kappa =\langle k \rangle/(N-1) $, are given by \cite{NAVPH}:
	\begin{align}	\label{Eq10a}	
	& \bar k(\varepsilon)=  \frac{N-1}{2\sinh(\alpha\beta\mu/2)} \Big ( e^{\alpha \beta \mu/2 }{}_{2}F_{1} \big (1, \alpha;  1+\alpha  ;-e^{ \beta \varepsilon } \big ) \nonumber \\
	 & -   e^{-\alpha \beta \mu /2}{}_{2}F_{1} \big (1, \alpha ;  1+\alpha  ;- e^{ \beta(\varepsilon -\mu)} \big ) \Big ), \\
&\kappa=  \frac{1 }{4\sinh^2(\alpha\beta\mu/2)} \Big ( e^{\alpha \beta \mu} {}_{3}F_{2} \big (1, \alpha, \alpha ; 1+\alpha , 1+\alpha  ;- e^{ \beta \mu } \big ) \nonumber \\
	& - 2 {}_{3}F_{2} \big (1, \alpha, \alpha ; 1+\alpha , 1+\alpha  ;- 1\big ) \nonumber \\	&+ e^{ -\alpha \beta \mu} {}_{3}F_{2} \big (1, \alpha, \alpha ; 1+\alpha , 1+\alpha  ;- e^{ -\beta \mu} \big )\Big ), 	\label{Eq10b}
	 \end{align}
where ${}_{p}F_{q}(a_1, \dots, a_p; b_1, \dots, b_q; z)$ is the generalized hypergeometric function \cite{AEWM,NIST}. 

Using the asymptotic properties of the generalized hypergeometric function and relation,
\begin{align}
		 {}_{3}F_{2} \big (1, a, a ; 1+a , 1+a  ;z\big ) =  - a^2\frac{\partial}{\partial a}\Big(  \frac{1}{a}\, {}_{2}F_{1} \big (1, a ; 1+a ;z\big ) \Big),
		 \label{A5}
	\end{align}	
in the limit of $T\ll T_c$ we obtain 
\begin{align}
\kappa =  \frac{\delta  + e^{ -\delta} -1}{4\sinh^2(\delta/2)} + \mathcal O(\alpha^2),	
\label{Eq13}
\end{align}
where $\delta = \beta_c(\gamma -1)\mu_0$ and $\mu_0 = \mu(0)$.

{\em Clustering coefficients.} --  For a given node $i$ with the energy $\varepsilon_i$, the local clustering coefficient, $ c(\varepsilon_i) $, can be calculated as follows \cite{BMPSR}:
\begin{align}
c(\varepsilon_i ) = \frac{ \sum_{j,k} p(\varepsilon_i, \varepsilon_j ) p(\varepsilon_j, \varepsilon_k  ) p(\varepsilon_k, \varepsilon_i  )}{\big(\sum_j p(\varepsilon_i, \varepsilon_j ) \big)^2}.
\end{align}
 Using this result, one can write the first GCC as
\begin{align}
  C_1 =&\frac{1}{N}\sum_i \frac{\sum_{j,k} p(\varepsilon_i, \varepsilon_j ) p(\varepsilon_j, \varepsilon_k  ) p(\varepsilon_k, \varepsilon_i  )}{(\sum_j  \bar k(\varepsilon_j))^2}.
    \label{C1a}
    \end{align}
The second GCC can be written as \cite{BBOR}:
\begin{align}
    C_2=&\frac{\sum_i \bar k(\varepsilon_i)(\bar k(\varepsilon_i) -1) c(\varepsilon_i ) }{\sum_i  \bar k(\varepsilon_i)(\bar k(\varepsilon_i) -1)}.
    \label{C2a}
    \end{align} 
    In the continuous limit, we obtain 
\begin{align}   \label{Cl} 
c(\varepsilon)= & \frac{\iint  p(\varepsilon, \varepsilon'  ) p(\varepsilon', \varepsilon''  ) p(\varepsilon, \varepsilon''  )\rho(\varepsilon' )\rho(\varepsilon'')  d \varepsilon' d \varepsilon'' }{\big (\int p(\varepsilon, \varepsilon' ) \rho(\varepsilon' ) d \varepsilon' \big )^2  }, \\
\label{C1h}
    C_1= &\int d\varepsilon\rho(\varepsilon)  \frac{\iint  p(\varepsilon, \varepsilon'  ) p(\varepsilon', \varepsilon''  ) p(\varepsilon, \varepsilon''  )\rho(\varepsilon' )\rho(\varepsilon'')  d \varepsilon' d \varepsilon'' }{\big (\int p(\varepsilon, \varepsilon' ) \rho(\varepsilon' ) d \varepsilon' \big )^2  }, \\
  C_2= &\frac{\iiint  p(\varepsilon, \varepsilon'  ) p(\varepsilon', \varepsilon''  ) p(\varepsilon, \varepsilon''  )\rho(\varepsilon) \rho(\varepsilon' )\rho(\varepsilon'') d \varepsilon d \varepsilon' d \varepsilon'' }{\int \big (\int p(\varepsilon, \varepsilon' ) \rho(\varepsilon' ) d \varepsilon' \big )^2 \rho(\varepsilon )d \varepsilon }.
    \label{C2b}
\end{align}

{\em Results.} --  We find that in a low-temperature regime, the clustering coefficients behave as
\begin{align} \label{C1}
	 C_1&=	1-  \frac{1}{8\sinh \delta}  +{\mathcal O}(\alpha ^2), \\
 	C_2 &=\frac{\delta \coth \delta}{2\sinh \delta} + {\mathcal O}(\alpha^2) ,
 	 \label{C2}
\end{align}
where $\delta =\mu_ 0\beta_c(\gamma -1)  /2$ and $\mu_0 = \mu(0)$.

\begin{figure}[tbh]
\includegraphics[width=1\linewidth]{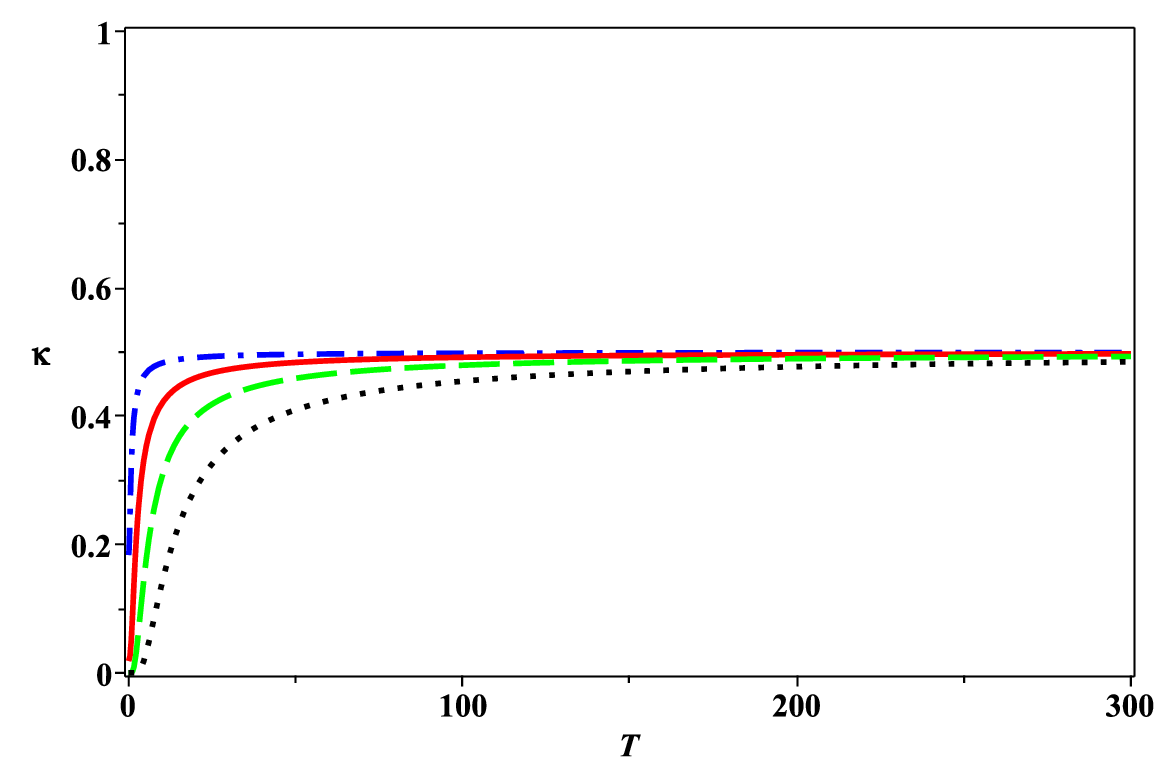} 
            	\caption{The average node degree per node, $ \kappa $, as a function of the temperature ($\gamma =2.1$, $T_c =1$). Black dotted line: $\mu =20$, green dashed line: $\mu =10$, red solid line: $\mu =5$. The blue dash-dotted line: $\mu =2$.  }
            	\label{fig3}
            \end{figure}

\begin{figure}[tbh]
\includegraphics[width=1.1\linewidth]{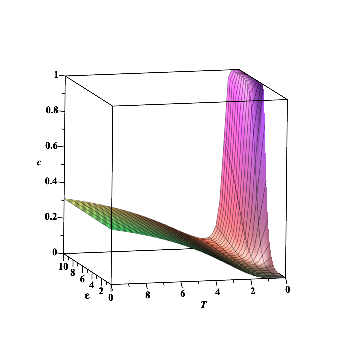} 
	\caption{The local clustering coefficient, $c$, as a function of the energy $\varepsilon$ and temperature $T$ ($\gamma =2.1$, $T_c =1$, $\mu =10$).  }
	\label{fig3d}
\end{figure}

In Figs.  \ref{fig3} -- \ref{fig3d}, the results of numerical simulations are presented. For illustrative purposes, we consider a model with temperature-independent chemical potential. Outcomes in Fig. \ref{fig3} confirm our analytical predictions for the behavior of average node degree per node (for details, see Ref. \cite{NAVPH}). As one can see, $\kappa \rightarrow 1/2$ for $T \gg T_c$. The behavior of the average node degree near zero temperature is described by Eq. \eqref{Eq13}. 

  \begin{figure}[tbh]
\includegraphics[width=1\linewidth]{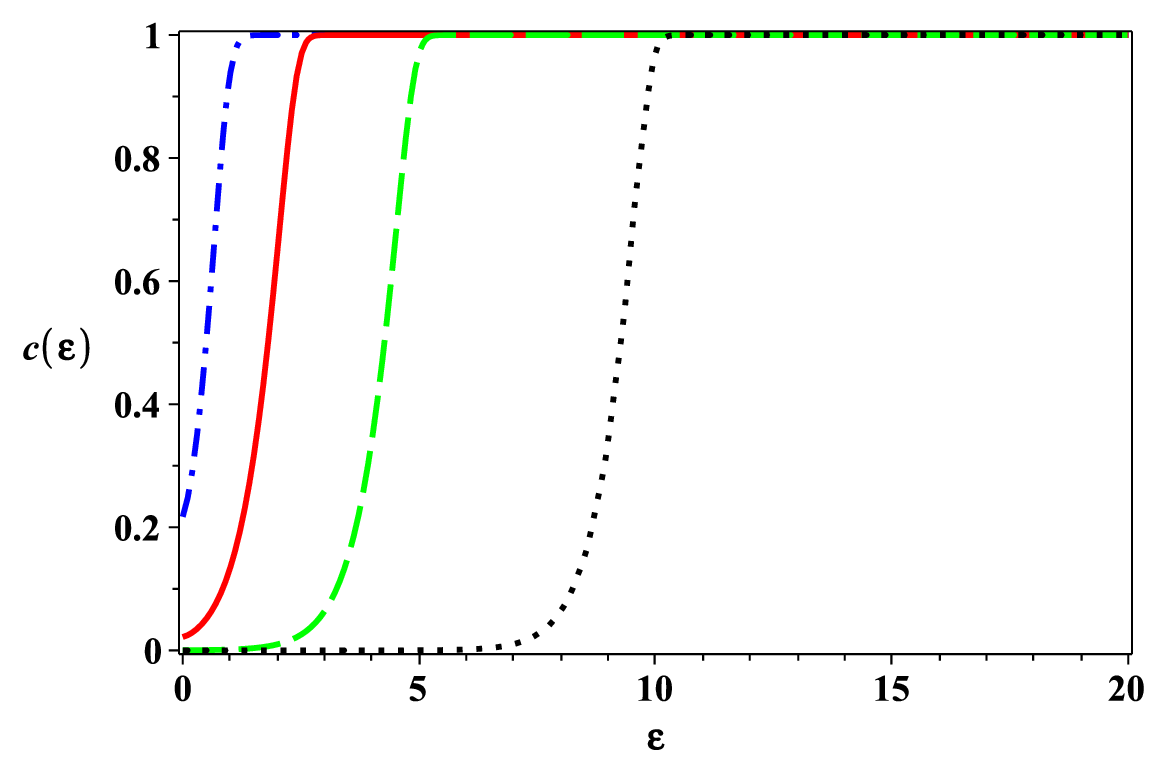} 
	\caption{The local clustering coefficient, $c(\varepsilon)$, as a function of the energy $\varepsilon$ ($\gamma =2.1$, $T_c =1$, $T =0.1$). Black dotted line: $\mu =20$, green dashed line: $\mu =10$, red solid line: $\mu =5$, blue dash-dotted line: $\mu =2$.  }
	\label{fig1e}
\end{figure}
                        
  \begin{figure}[tbh]
\includegraphics[width=1\linewidth]{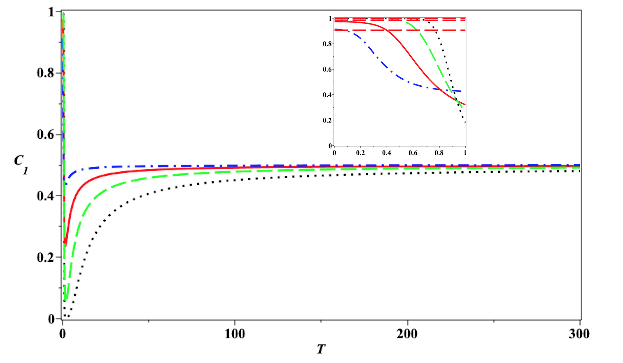} 
	\caption{The global clustering coefficient, $C_1$, as a function of the temperature ($\gamma =2.1$, $T_c =1$). Black dotted line: $\mu =20$, green dashed line: $\mu =10$, red solid line: $\mu =5$, blue dash-dotted line: $\mu =2$. The inset is a zoom of the main figure with the same lines convention. The dash-dotted red lines present the asymptotic value of the clustering coefficient as $T \rightarrow 0$ (see Eq. \eqref{C1}). }
	\label{fig3c}
\end{figure}

  \begin{figure}[tbh]
\includegraphics[width=1\linewidth]{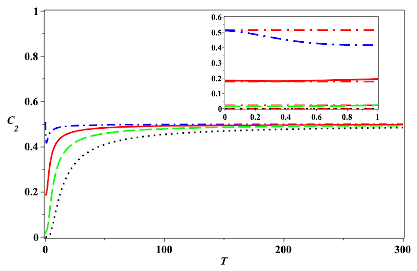} 
	\caption{The global clustering coefficient, $C_2$, as a function of the temperature ($\gamma =2.1$, $T_c =1$). Black dotted line: $\mu =20$, green dashed line: $\mu =10$, red solid line: $\mu =5$, blue dash-dotted line: $\mu =2$. The inset is a zoom of the main figure with the same lines convention. The dash-dotted red lines present the asymptotic value of the clustering coefficient as $T \rightarrow 0$ (see Eq. \eqref{C2}). }
	\label{fig2c}
\end{figure}

In Fig. \ref{fig3d}, the LCC is represented as a function of the energy and temperature for the fixed value of the chemical potential ($\mu = 10$). In Fig. \ref{fig1e}, the LCC is depicted for the fixed temperature ($T=0.1$) and different magnitudes of the chemical potential: $\mu = 2,5,10,20$. As one can observe, the LCC tends to its maximum value for $T \ll T_c$ and high energies. It indicates the tendency of nodes with high energy to have clustered neighbors.

In Figs. \ref{fig3c},\ref{fig2c}, the global clustering coefficients, $C_1$ and $C_2$, are depicted for different magnitudes of the chemical potential. Both clustering coefficients behave according to the theoretical predictions for high temperatures,  {$C_{i} \rightarrow 1/2$} ($i=1,2$) when $T \rightarrow \infty$. However, for low temperatures, the results are quite different. As follows from Fig. \ref{fig3}, the network becomes sparse in the low temperatures limit, so that $\kappa \ll 1$ when  $T\rightarrow 0$. Therefore, one expects the clustering coefficient to behave similarly,  $ C_i\ll 1$ when  $T\rightarrow 0$. While the first definition yields the `wrong' result, the behavior of $C_{2}$ is in agreement with the behavior of the average node degree (see Fig. \ref{fig3}).   

The difference between the predictions can be explained as follows. The GCC $C_1$ captures the average tendency of nodes to have clustered neighbors and shows how ``clumpy" the network is on average, even if there aren't many complete triangles. The  GCC $C_2$ describes the presence of complete ``closed" triangles, representing tightly knit communities. A high value implies a network with many tightly clustered groups, while a low value suggests a more random or dispersed structure. Our findings show that in the low-temperature regime, the network exhibits a high average tendency of nodes to have clustered neighbors, even if they don't necessarily form complete triangles, and to form a dispersed structure. 
 
 In summary, we have shown that the global clustering coefficients $C_1$ and $C_2$ yield different predictions in the low-temperature regime. Since the global clustering coefficient's varied behaviors can significantly affect the accuracy outcome, great care must be taken when choosing it for a given application.    

The author acknowledges the support of the CONAHCYT.

%


\end{document}